\documentclass[twocolumn, pra,showpacs,superscriptaddress]{revtex4}
\usepackage{graphicx}
\usepackage{dcolumn}
\usepackage{bm}
\usepackage{amsmath}

\begin{document}

\title{The properties of Tonk-Girardeau Gas at Finite Temperature and Comparison with Polarized Free Fermions}
\author{Yajiang Hao}
\email{haoyj@ustb.edu.cn}
\affiliation{Department of Physics, University of Science and Technology Beijing, Beijing 100083, China}
\author{Yafei Song}
\affiliation{Department of Physics, University of Science and Technology Beijing, Beijing 100083, China}
\author{Xiaochen Fu}
\affiliation{Department of Physics, University of Science and Technology Beijing, Beijing 100083, China}

\date{\today}

\begin{abstract}
In the present paper we investigate the Tonks-Girardeau gas confined in a harmonic trap at finite temperature with thermal Bose-Fermi mapping method. The pair distribution, density distribution, reduced one-body density matrix, the occupations number of natural orbitals, and momentum distribution are evaluated. In the whole temperature regime the pair distribution and density distribution exhibit the same properties as those of polarized free Fermions because both of them depend on the modulus of wavefunction rather than wavefunction. While the reduced one-body density matrix, the natural orbital occupation, momentum distribution, which depend on wavefunction, of Tonks gas displays Bose properties different from polarized free Fermions at low temperature. At high temperature we can not distinguish Tonks gas from the polarized free Fermi gas by all properties qualitatively.
\end{abstract}


\maketitle



\section{introduction}

For the high controllability and 'purity' of quantum gas, it has been the popular platform to investigate novel quantum phenomena since it was realized in experiment. The cold atoms can be frozen to the zero point oscillation in transverse with highly anisotropic magnetic trap or two dimensional optical lattices \cite{Ketterler,Paredes,Toshiya}, and thus the cold atoms become a one-dimensional (1D) quantum gas of strong correlation in longitudinal direction \cite{Olshanii}. The interactions between atoms will be the effective 1D contact interaction depending on the $s$-wave scattering length and its strength can be tuned from the weakly interacting regime to the strongly interacting regime with Feshbach resonance technique and confinement induced resonance. In this way the Bose gas of strong repulsive interaction, Tonks-Girardeau (TG) gas \cite{GirardeauJMP,Tonks}, can be realized, and offer researchers an opportunity to investigate the quantum many body physics of  previously so-called 'toy' model. Besides the realization of arrays of 1D gases, Jacqmin etc.'s experiment realized a single 1D Bose gas close to the strongly interacting regime and the atom number fluctuation measurements can be performed in small slices of gases \cite{Jacqmin}. This allow us to study the properties of 1D Bose gas related with temperature and the interplay effect between interaction and temperature.

Theoretically the unconfined 1D Bose gas of contact interaction corresponds to the Lieb-Liniger model and can be exactly solved with Bethe-ansatz method in the whole interacting regime at zero temperature and at finite temperature \cite{LiebLiniger,CNYang}. Combining the exact solution of integral model with local density approximation we can study the 1D quantum gas confined in a harmonic trap \cite{GaoXL, Hao09,Yin2009,GuanXWPRL}. The exact wave function of strongly interacting TG gas in a harmonic trap can be obtained with Bose-Fermi mapping method \cite{ALenard,GirardeauJMP,Girardeau02}. So far most previous research focus on the ground state properties of 1D Bose gas. Its properties at finite temperature were not the focus of research although the temperature effect is important. The investigation of ground state properties show that as the atomic interaction becomes strong the ground state density profiles of Bose gas evolve from the Bose properties to the ``fermionized" characters same as the polarized free Fermi gas \cite{Hao06,Hao09,Zoellner,Deuretzbacher}. At zero temperature, the equilibrium state of 1D Bose gas is coherent quasicondensate in weakly interaction regime, and the gas reaches fermonized TG regime in strongly interacting regime. Although the TG gas show the same density profiles as free Fermi gas, the density matrix and the momentum distribution exhibit the typical Bosonic properties \cite{ALenard,Hao07}.

Based on two-body correlation the interacting 1D Bose gas can be classified as various physical regimes with the change of temperature and interaction \cite{ShlyapnikovPRL,ShlyapnikovPRA,Jacqmin,PDeuar}, and the interplay between interaction and temperature is an important topic. At low temperature the 1D Bose gas might be in the strongly interacting TG regime, a coherent regime or a fully decoherent quantum regime. Although the finite-temperature effect cannot be discounted, the theoretical determination of correlations, energy, and momentum scales at finite temperature has only been partially investigated \cite{ShlyapnikovPRL,ShlyapnikovPRA,JBrand,PDeuar,MKormos,HHu,ShlyapnikovDisorder,AVogler,MPanfil,PVignolo,MDHoffman,GLang,MRigol}. In the present paper we will investigate the TG gas trapped in a harmonic trap at finite-temperature. With thermal Bose-Fermi mapping method the pair distribution function, the reduced one-body density matrix (ROBDM), the occupation of natural orbitals, and momentum distributions are evaluated for different temperature. As a comparison the corresponding result of polarized free Fermi gas are also exhibited.

The paper is organized as follows. In Sec. II, we give a brief review of the model of 1D TG gas and method. In Sec. III, we present the pair distribution function, ROBDM, the occupation of natural orbitals, and momentum distributions of TG gases for different temperatures. A brief summary is given in Sec. IV.

\section{Model and method}

In Bose quantum gases the properties of system are governed by the contact interaction between atoms, the strength of which depends on the $s$-wave scattering length $a_s$. When the cold atoms are confined intensively in transverse direction with tapping frequency $\omega_{\bot}$, the quantum gas becomes an effective one-dimensional many body quantum system. The effective 1D interaction strength between atoms is
\begin{equation}
g_{1D}=\frac{4\hbar^2a_s^2}{ma_{\bot}}(1-\mathcal{C}\frac{a_s}{a_{\bot}})^{-1}							\label{g1D}
\end{equation}
with $m$ being atom mass, the transversal length unit $a_{\bot}=\sqrt{\hbar/m\omega_{\bot}}$, and dimensionless constant $\mathcal{C}=1.4603$.

For the 1D quantum gas of $N$ Bose atoms confined in a one-dimensional harmonic trap $V(x)=\frac12 m\omega^2x^2$ with $\omega$ being the trapping frequency, the system can be described by the Hamiltonian
\begin{equation}
H=-\sum_{i=1}^N\left [\frac{\hbar^2}{2m}\frac{\partial^2}{\partial x_i^2}+V(x_i)\right ] +g_{1D}\sum_{1\leq i < j\leq N}\delta(x_i-x_j),  \label{Hamiltonian}
\end{equation}
where $x_i$ is the position of $i$th atom. Experimentally the effective 1D interacting constant $g_{1D}$ can be tuned by Feshbach resonance or confinement induced resonance in the full interacting regime from strongly attractive limit to strongly repulsive limit.

In the present paper we shall focus on the strongly repulsive interaction limit, i.e., the TG gas. In this case the many body wavefunction $\psi^b(x_1,x_2,...,x_N)$ is the solution of
\begin{equation}
H\psi^b(x_1,x_2,...,x_N)=E\psi^b(x_1,x_2,...,x_N)		  \label{EigenEq}
\end{equation}
with $H=-\sum_{i=1}^N\left [\frac{\hbar^2}{2m}\frac{\partial^2}{\partial x_i^2}+\frac12m\omega^2x_i^2\right ]$.
For the strong interaction between atoms the wavefunction should  satisfy the constraint
\begin{equation}
\psi^b(x_1,...,x_i,...,x_j,...,x_N)=0		\label{BC}
\end{equation}
if $x_i=x_j$ ($1\leq i < j \leq N$).
By Bose-Fermi mapping method \cite{ALenard,Girardeau02} to all the eigen wavefunction $\psi^b(x_1,x_2,...,x_N)$ of eigen equation Eq.(\ref{EigenEq}) of Bose system there are corresponding wavefunction of polarized free fermions $\psi^{f}(x_{1},x_{2},...,x_{N})$
\begin{equation}
\psi ^{b}(x_{1},x_{2},...,x_{N})=A(x_{1},x_{2},...,x_{N})\psi^{f}(x_{1},x_{2},...,x_{N}),  \label{BFM}
\end{equation}
where the function $A$ is defined by $A(x_{1},x_{2},...,x_{N})=\prod\limits_{1\leq i\leq j\leq N}sign(x_{i}-x_{j})$ and $sign(x)$ is +1, 0, -1 for $x$ being positive, zero, and negative, respectively. Since the wavefunction of the polarized free Fermions is antisymmetry under exchanges, the constraint Eq. (\ref{BC}) can be satisfied automatically. It is well known the many body wavefunction of free Fermions can be formulated as a Slater determinants of single particle eigen function $\phi_{\nu}(x)$
\begin{equation}
\psi^f(x_1,x_2,...,x_N)= \frac1{\sqrt{N!}}\det_{i,j=1}^N \phi_{\nu_i}(x_j). \label{WFF}
\end{equation}
The eigen function of harmonic oscillator take the form of $\phi _{\nu }(x)=\frac {H_{\nu }(x)\exp (-x^{2}/2)}{\pi ^{1/4}\sqrt{2^{\nu }\nu !}} $ with $H_{\nu }(x)$ Hermite polynomial. As the set $\alpha=\{\nu_1,\nu_2,...,\nu_N\}$ take different combination, we can have all eigen wavefunctions including the ground state and excited states, which satisfies the eigen equation $H\psi^b_{N,\alpha}(x_1,x_2,...,x_N)=E_{N,\alpha}\psi^b_{N,\alpha}(x_1,x_2,...,x_N)$, where the eigen energy $E_{N,\alpha}=\sum_{i=1}^N(\nu_i+1/2)\hbar \omega$.

At zero temperature the system can be described by the ground state with $\alpha=\{0, 1, \dots, N-1\}$, while at finite temperature the TG gases are governed by statistical mechanical ensemble of Bosons. The TG gases are described by the probability $P_{N,\alpha}$ attached to eigenstate $\psi^b_{N,\alpha}(x_1,x_2,...,x_N)$, which take the form of $P_{N,\alpha }=Z^{-1} \prod \limits_{l=1}^{N}e^{-\beta (\varepsilon _{\nu _{l}}-\mu )}$ with $Z$ being the partition function, $\beta=1/k_BT$, the single particle energy $\varepsilon_{\nu_l}=(\nu_l+1/2)\hbar\omega$, and $\mu$ chemical potential.

By the Bose-Fermi transformation Eq. (\ref{BFM}) the eigen wavefunction of TG gas distinguish from the corresponding wavefunction of polarized free Fermions by the signature function $A(x_{1},x_{2},...,x_{N})$. Therefore the properties that depend on the modulus of wavefunction such as the pair distribution and single particle density distribution are same for TG gases and Fermi gases, and the properties that depend on wavefunction such as ROBDM and momentum distribution are distinct with each other.

Physically the pair distribution can be seen as the joint probability density that in two successive measurement if the first measurement finds one atom is at $x_1$, then an immediate second measurement finds an atom at $x_2$. At finite temperature the pair distribution function of TG gases and free Fermion gases are same, $D_{2}^{b}(x_1,x_2)=D_{2}^{F}(x_1,x_2)$, which take the form of
\begin{eqnarray}
D_{2}^{b}\left(  x_{1},x_{2}\right) =\sum_{N=2,\alpha}^{\infty}P_{N,\alpha}\frac{N!}{(N-2)!}\int_{\Omega}dx_{3}...\int_{\Omega}dx_{N}\psi_{N,\alpha}^{f}\psi_{N,\alpha}^{f\ast}.	\nonumber
\end{eqnarray}
Inserting the wavefunction into the above equation we can obtain the following simplified formula
\begin{align}
D_{2}^{b}(x_{1},x_{2})  & =\det_{1\leq i,l\leq2}\rho_{1F}\left(  x_{i},x_{l}\right)  \\
& =\sum_{\{\nu_1,\nu_2\}}f_{\nu_{1}}f_{\nu_{2}}\sum_{P}(-1)^{P}{\displaystyle\prod\limits_{i=1}^{2}}
\phi_{\nu_{i}}(x_{i})\phi_{\nu_{P(i)}}^{\ast}(x_{i}),			\nonumber
\end{align}
where $\{\nu_1,\nu_2\}$ is the two dimension permutations of energy level index $\{0,1,\dots,M\}$ with $M$ being the highest single particle energy level to be considered. In the calculation, we omit those high energy levels whose corresponding Fermi-Dirac distribution $f_{\nu}=e^{-\beta (\varepsilon _{\nu }-\mu )}$ is negligible small.

The ROBDM can be viewed as the probability to find the particles at positions $x$ and $y$ in two successive measurements, respectively. At finite temperature the ROBDM $\rho ^{b}(x,y)$ is defined as \cite{ALenard}
\begin{eqnarray}
\rho ^{b}(x,y) =\sum_{N=1}^{\infty }\sum_{\alpha }P_{N,\alpha }\frac{N!}{(N-n)!}
\int_{-\infty }^{\infty}dx_{2}...\int_{-\infty }^{\infty}dx_{N} \nonumber \\
\times \psi _{N,\alpha }^{b}(x,x_{2},...,x_{N})
\psi _{N,\alpha }^{b\ast }(y,x_{2},...,x_{N}).    \label{ROBDM1}
\end{eqnarray}
Here $\alpha$ sums over all possible permutations of $N$ dimension of $\{0,1,...,M\}$. In the calculations we also ignore the eigenstates with negligible $f_{\nu}$. With Eq. (\ref{BFM}) the Boson ROBDM $\rho^b(x,y)$ is expressed in term of the determinants of one-body Fermion density matrix $\rho_{1F}(x,y)$ as follows \cite{ALenard,PVignolo}
\begin{eqnarray}
\rho ^{b}\left( x,y\right) =\sum_{j=0}^{\infty }\frac{(-2)^{j}}{j!}(sign(x-y))^{j}\int_{y}^{x}dx_{2}...dx_{j+1} \nonumber \\
\times \det_{1\leq n,m\leq j+1}\rho^f\left( x_{n},x_{m}^{\prime }\right).  \label{ROBDM2}
\end{eqnarray}
With the properties of determinants $ \det_{n,m}\rho^f(x_{n},x_{m}^{\prime })$ take the form of  \cite{PVignolo}
\begin{eqnarray}
&&  \det_{1\leq n,m\leq j+1}\rho ^f\left( x_{n},x_{m}^{\prime }\right)    \label{Det1F}   \\
&=&\sum_{\nu _{1}...\nu _{j+1}}f_{\nu _{1}}...f_{\nu_{N}}\sum_{P}(-1)^{P}\prod\limits_{i=1}^{j+1}\phi_{\nu _{i}}\left(x_{i}\right) \phi_{\nu _{P(i)}}^{\ast }\left( x_{i}\right). \nonumber
\end{eqnarray}
Thus the multiple integral in Eq. (\ref{ROBDM2}) can be reduced into single-variable integral, which makes the evaluation of ROBDM become possible. In the calculation more high single particle orbitals should be included in the summation as temperature increases, so the computational amount of the exact implementation of Eq. (\ref{Det1F}) will beyond our ability. For high temperature we will follow the approximation of the one-body Fermion density matrix used in Eq. (10) in Ref. \cite{PVignolo}.

After we have reduced one-body density matrix $\rho(x,y)$, which might be ROBDM of TG gas $\rho^b(x,y)$ or that of Fermions $\rho^f(x,y)$, by solving the following eigen equation of $\rho(x,y)$
\begin{equation}
\int_{-\infty}^{\infty}\rho \left( x,y\right)\varphi_i(y)dy =\lambda_i\varphi_i(x)  \label{NO}
\end{equation}
we can obtain the natural orbital $\varphi_i(x)$ with $\lambda_i$ being the occupation number of natural orbital, which satisfies $\sum_i\lambda_i=N$. The momentum distribution can be evaluated by the Fourier transform of ROBDM
\begin{equation}
n(k)=(2\pi)^{-1}\int_{-\infty}^{\infty}dxdy\rho\left( x,y\right) e^{-ik(x-y)}.  \label{MD}
\end{equation}

\section{Numerical Results}

\begin{figure}[tbp]
\includegraphics[width=3.4in]{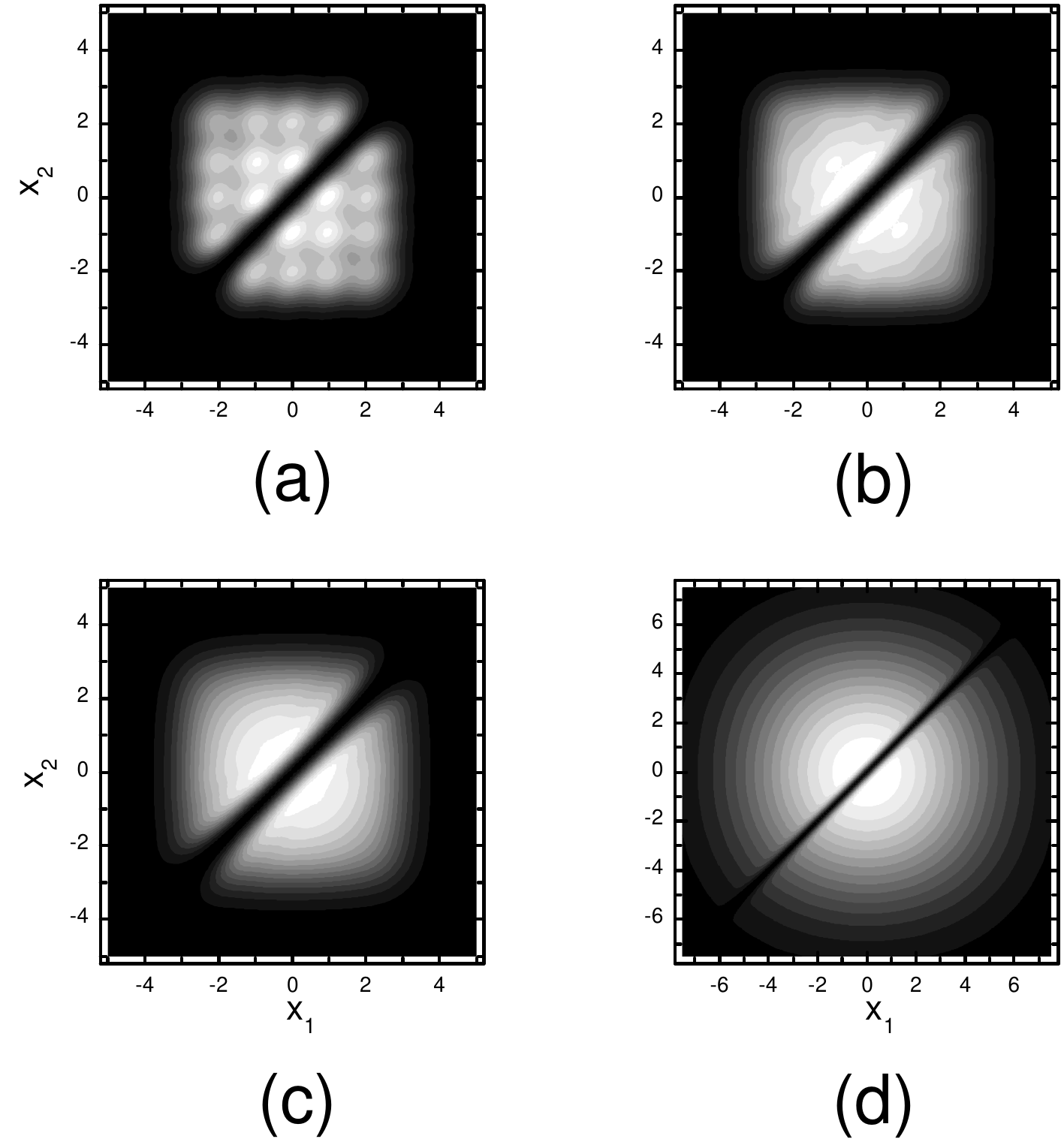}\newline
\caption{(color online) Pair distribution for 5 Bosons and Fermions in a harmonic trap, which is same with each other. (a) $T=0.1\hbar\omega$; (b) $T=0.5\hbar\omega$; (c) $T=1.0\hbar\omega$; (d) $T=10.0\hbar\omega$.} \label{PD}
\end{figure}
We will present the pair distributions, ROBDM, occupation number of natural orbitals and momentum distributions of 1D TG gases in a harmonic trap with $N=5$. As a comparison, we also show the results of polarized free Fermions. For simplicity in this paper we take $\sqrt{\hbar/m\omega}$ and $\hbar\omega$ as the length unit and temperature unit.

The pair distribution function $D_2(x_1,x_2)$ depends on the absolute value of the wavefunction, so it is the same for the bosons and fermions. In Fig. \ref{PD} the pair distribution function of TG gas with $N=5$ are plotted at different temperatures. At low temperature (for example, $T=0.1$) the pair distribution function exhibit the same properties as those of TG gases at zero temperature and the temperature effect is not obvious \cite{Girardeau2001}. The diagonal part (the region close to $x_1=x_2$) means the probability that two Bosons appear at the same position, which is zero because the interactions between bosons are strong and two atoms separate as much as possible. With the increase of $|x_1-x_2|$, the joint probability density increases with oscillation first and decrease to zero at large enough distances. As temperature becomes high although the joint probability density also show the behavior that increases first and then decreases, the oscillation vanish completely. In addition, with the increase of temperature the joint probability density $D_2(x_1,x_2)$ decreases at small distance of $|x_1-x_2|$ and  distributes in larger regions.


\begin{figure}[tbp]
\includegraphics[width=3.5in]{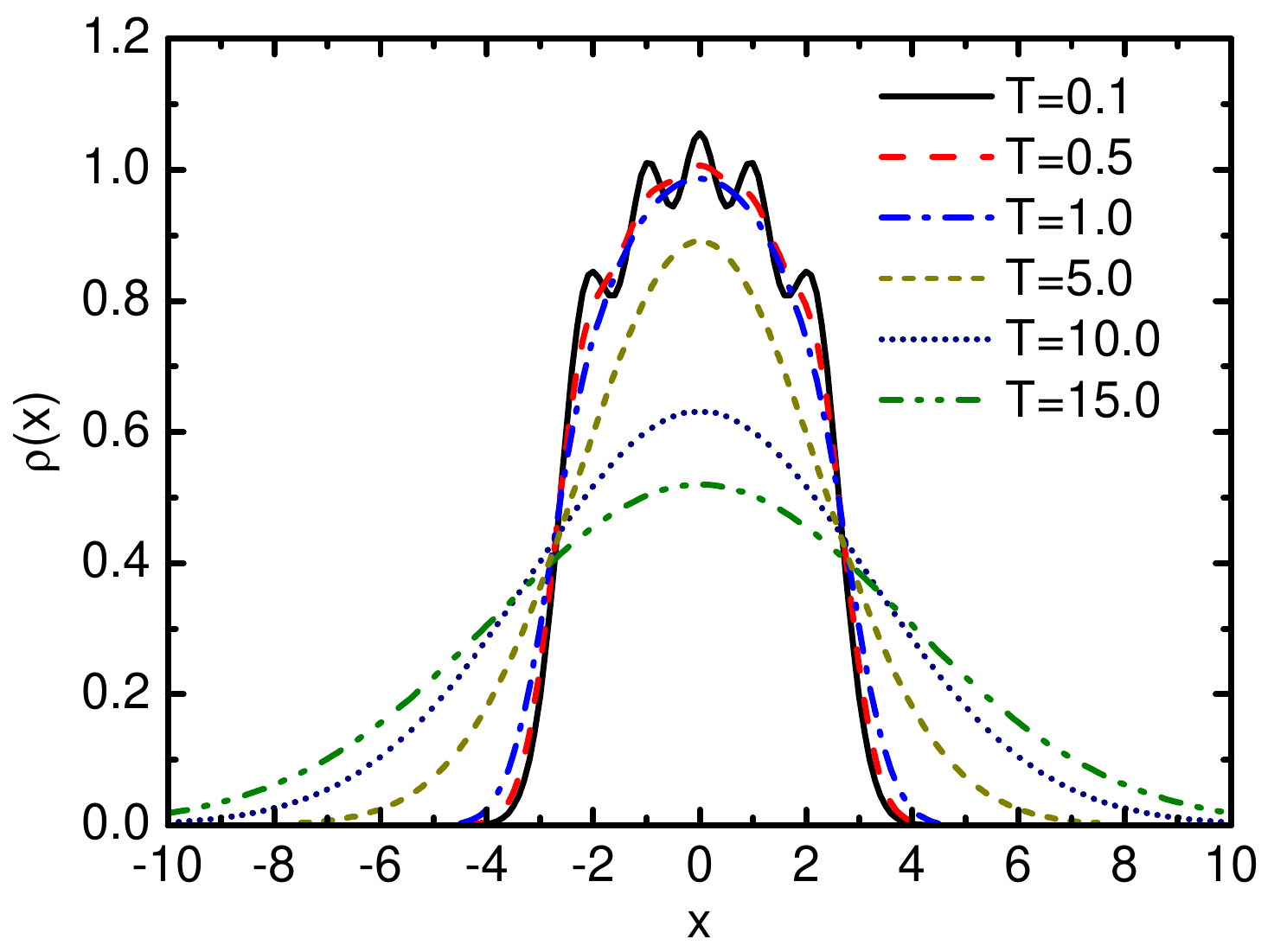}\newline
\caption{(color online) Density profiles of the ground state of TG gases in a harmonic trap at different temperatures with $N$=5. The temperature is in unit of $\hbar\omega$.} \label{density}
\end{figure}

In Fig. \ref{density} the density profiles of TG gases with $N=5$ at finite temperatures are plotted, which is the diagonal element of ROBDM $\rho(x)=\rho^b(x,x)$. At low temperature the density profiles display shell structure same as the case at zero temperature \cite{Hao06} and Bose atoms occupy in the center region of harmonic trap. In this case atoms populate at lower single particle energy levels with large probabilities.  As temperature become high the thermal energy of Bose atoms increases and the population probability at higher energy levels will increase. The shell structure of density profiles vanish and Bose atoms distribute in wider region. As temperature is high enough the property of Gauss distribution is exhibited. The density profiles can be fitted to the Gauss distribution $\rho(x)=\frac{N}{\sqrt{2\pi T}}\exp(-\frac12\frac{x^2}{T})$.

\begin{figure}[tbp]
\includegraphics[width=3.6in]{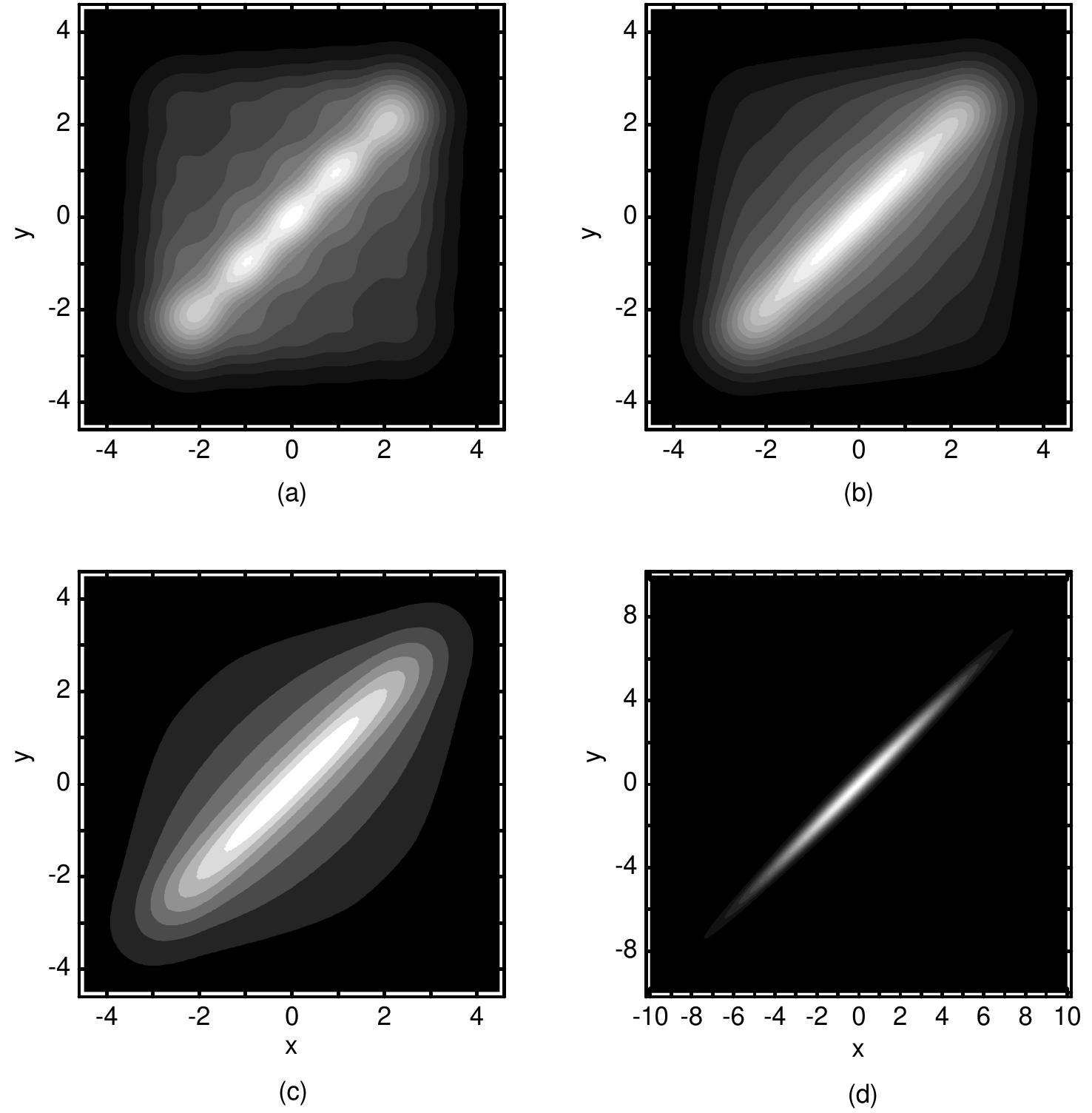}\newline
\caption{(color online) Reduced one-body density matrix of TG gases in a harmonic trap at different temperatures with $N$=5.  (a) $T=0.1\hbar\omega$; (b) $T=0.5\hbar\omega$; (c) $T=1.0\hbar\omega$; (d) $T=10.0\hbar\omega$.} \label{ROBDM}
\end{figure}

\begin{figure}[tbp]
\includegraphics[width=3.5in]{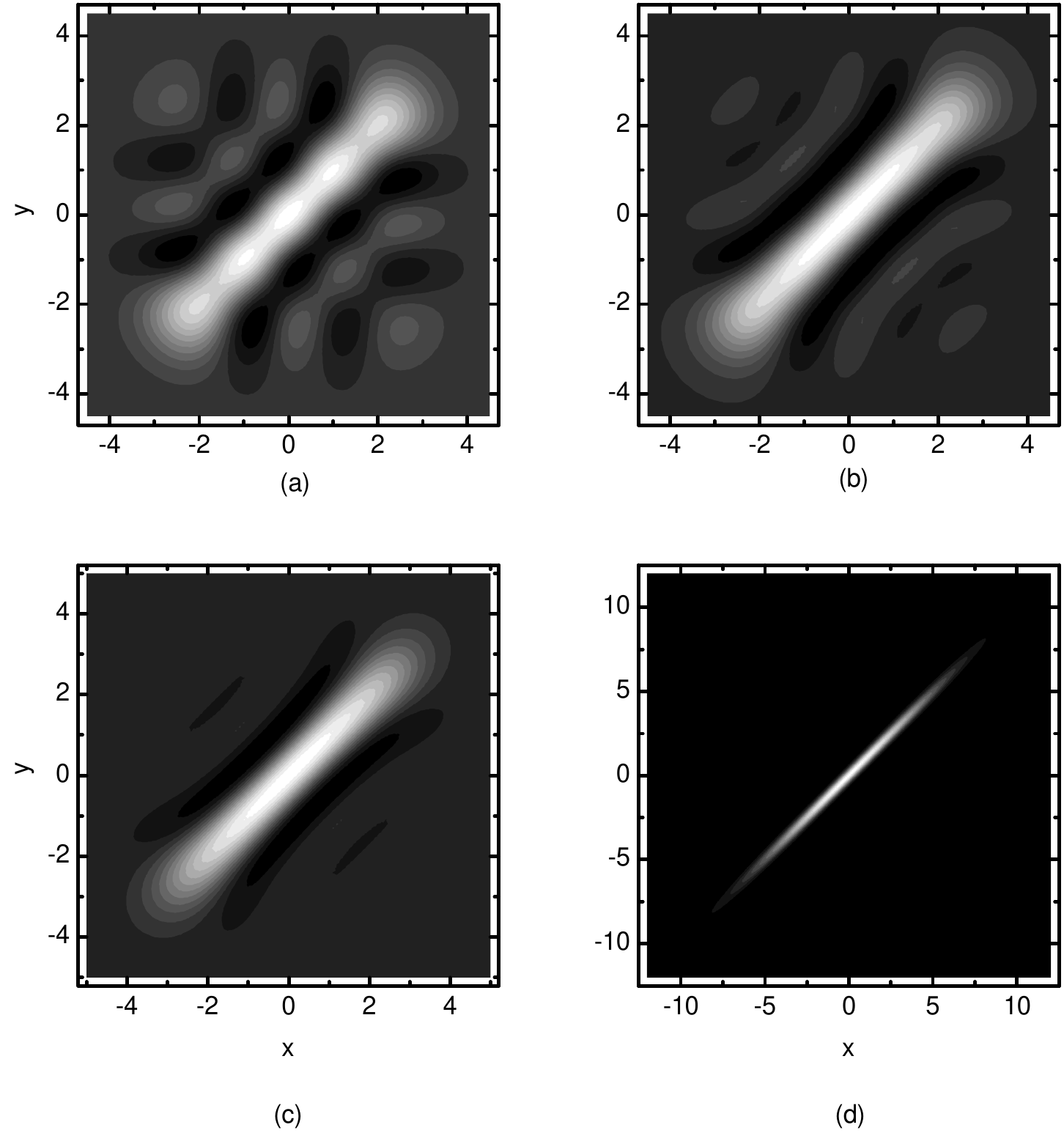}\newline
\caption{(color online) Reduced one-body density matrix of the ground state of polarized free fermions in a harmonic trap at different temperatures with $N$=5.  (a) $T=0.1\hbar\omega$; (b) $T=0.5\hbar\omega$; (c) $T=1.0\hbar\omega$; (d) $T=10.0\hbar\omega$.} \label{FermiROBDM}
\end{figure}

The reduced one-body density matrix of TG gases $\rho^b(x,y)$ with $N=5$ at finite temperature are shown in Fig. \ref{ROBDM}. At low temperature the ROBDM display the similar properties to those at zero temperature. Although the ROBDM is diagonal dominant, the off-diagonal elements are also not neglected. This is related to ODLRO and it is shown that there exists ODLRO in TG gases at low temperatures. With the increase of temperature $\rho^b(x,y)$ decreases more rapidly at large $|x-y|$ than the diagonal part. At high temperature the off-diagonal elements vanish completely and only the elements $\rho^b(x,y)$ for small $|x-y|$ have finite values. As a comparison we display the ROBDM of polarized free Fermi gas $\rho^f(x,y)$ with $N=5$ in Fig. \ref{FermiROBDM}. It is shown that ROBDM exhibit completely different behaviors from those of TG gases at low temperature. Even at very low temperature ($T=0.1\hbar\omega$) the off-diagonal element of ROBDM are negligibly small and there exists not ODLRO.  At high temperature the Fermi gases exhibit similar behaviors to TG gases. As the temperature is in the middle regime (for example, $T=1.0\hbar\omega$) the off-diagonal elements of $\rho^b(x,y)$ are much larger thatn those of $\rho^f(x,y)$.
\begin{figure}[tbp]
\includegraphics[width=3.5in]{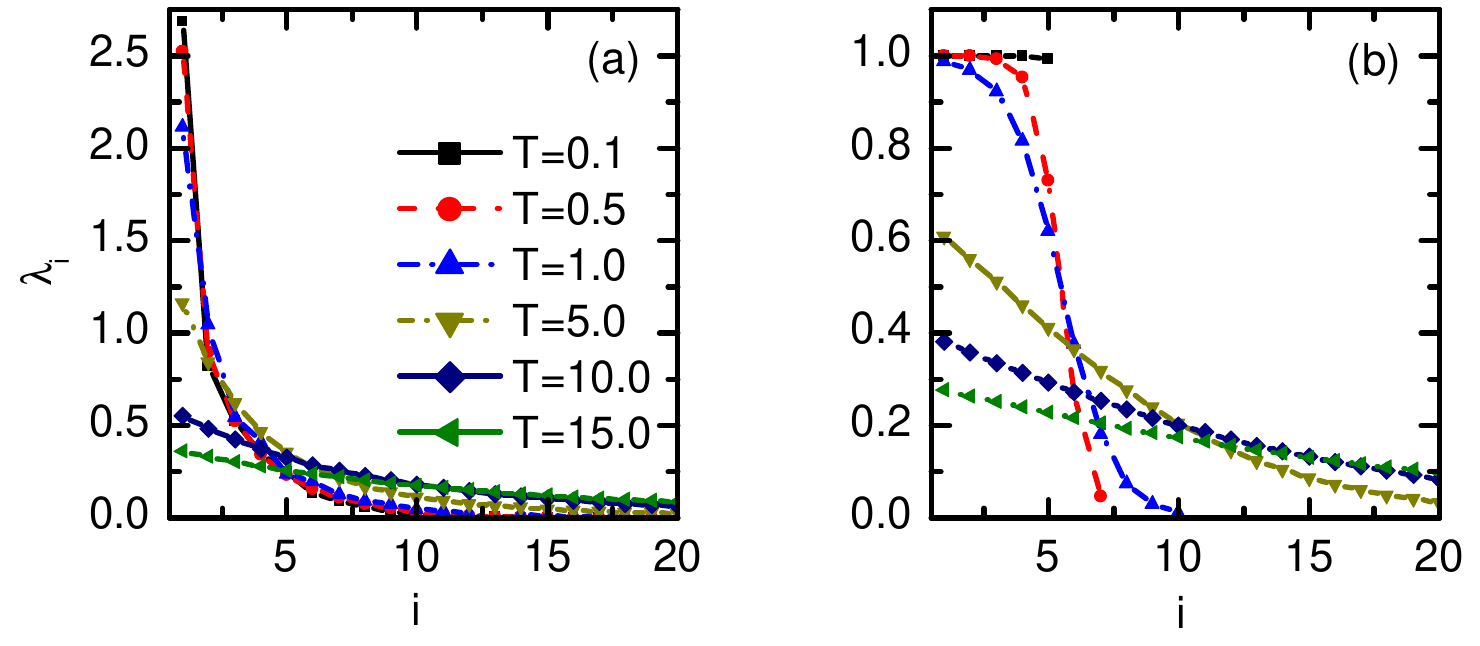}\newline
\caption{(color online) Occupation number of natural orbitals of TG gases (a) and polarized free Fermions (b) in a harmonic trap at different temperatures with $N$=5. The temperature is in unit of $\hbar\omega$.} \label{Ocu}
\end{figure}

The ODLRO is related to the occupation number of natural orbital $\lambda_i$ besides the behavior of ROBDM at large distance $|x-y|$. In macroscopic system if the occupation number $\lambda_0$ of the lowest natural orbital is much larger than one, the system will have ODLRO and exhibit BEC-like coherence effect. The system can be described by the associated natural orbital that play the role of an order parameter. In Fig. \ref{Ocu} the occupation of natural orbital of TG gases and polarized free fermions at different temperatures are displayed. For Fermi gas at $T=0.1\hbar\omega$ the occupation number is one for the $N$ lowest natural orbitals, while for higher natural orbitals the occupation numbers are zero. The distribution of occupations versus the orbital number $i$ behave as a step-like function. As the temperature increase more higher natural orbitals will be occupied and the occupation numbers of the lower natural orbitals decrease. The occupation distribution of TG gases at high temperature is similar to free fermions. But at low temperature the occupation of the lowest natural orbital is greatly larger than those of other naturals ($\lambda_0=2.69$ at $T=0.1\hbar\omega$). This means that the lowest natural orbital dominates the properties of the whole TG gases at low temperatures. With the increase of temperature the range of significantly occupied higher orbitals increase and the lowest natural orbitals shall not dominate the properties of TG gases, we have to consider more natural orbitals to describe the system.


\begin{figure}[tbp]
\includegraphics[width=3.5in]{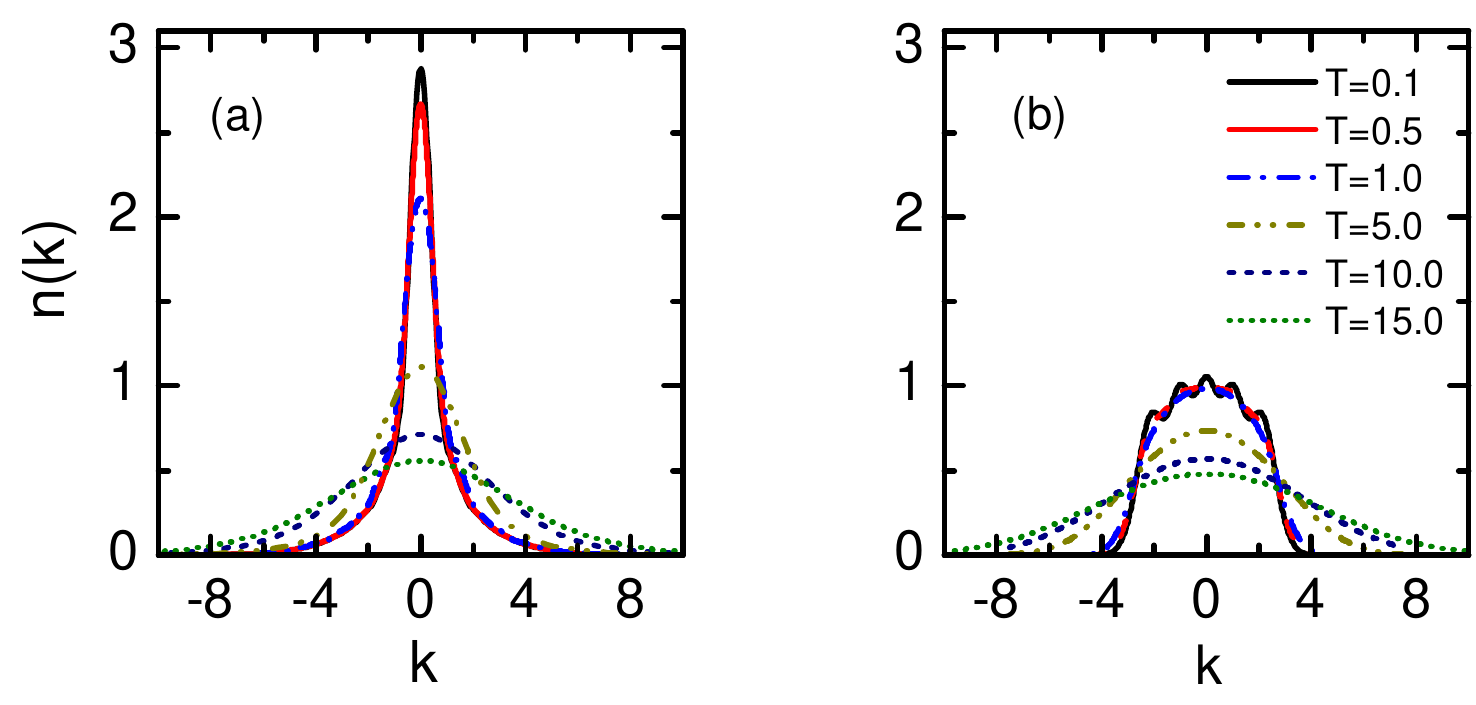}\newline
\caption{(color online) Momentum distribution of TG gases (a) and polarized free Fermions (b) in a harmonic trap at different temperatures with $N$=5. The temperature is in unit of $\hbar\omega$.} \label{Mom}
\end{figure}

In Fig. \ref{Mom}a the momentum distribution of TG gases with $N=5$ are displayed at finite temperature. At low temperature momentum distribution behave as a sharp peak structure at zero momentum point, which is the typical property of Bose system. The Bose atoms appear around the zero momentum region as the largest probability and the occupation probability decrease to zero rapidly at finite momentum regime. The height of peak shrinks with the increase of temperature and Bose atoms occupy large momentum regime because of the assistance of thermal energy resulting from the finite temperature. At high temperature the sharp peak structure of momentum distribution disappears and Bose atoms distribute in wider momentum regimes. As a comparison the momentum distributions of polarized free Fermions of $N=5$ are displayed in Fig. \ref{Mom}b. The momentum distributions in momentum space are exactly same as the density profiles in real space. It deserves to emphasize that the difference from those of TG gases mainly exhibit at low temperature. In this situation free Fermi gases show shell structure same as the distribution in the real space. This is the main difference between the Bose system and Fermi system.  Similar to the properties of ROBDM we also can not distinguish TG gases and polarized free Fermi gases by the momentum distributions at high temperature.

\section{conclusion}
In conclusion with thermal Bose-Fermi mapping method we calculated pair distribution function, density profiles, ROBDM, the occupation number of natural orbitals, and momentum distribution of strongly interacting TG gases at finite temperature.

In the whole temperature regime the diagonal part of pair distribution function $D_2(x_1,x_2)$ is zero, which shows that two strongly interacting atoms always separate as much as possible. Comparison to the case at low temperature $D_2(x_1,x_2)$ distribute in larger regimes at high temperature . At low temperature the density profiles exhibit the Fermion-like shell structure, and the shell structure vanish with the increase of temperature. At high temperature the density profiles can be fitted as Gaussian distribution $\rho(x)=\frac{N}{\sqrt{2\pi T}}\exp(-\frac12\frac{x^2}{T})$. The investigation of ROBDM shows that at low temperature there exist ODLRO in TG gas similar to the properties of zero temperature case, while the ODLRO vanish at high temperature. Correspondingly, at low temperature momentum distribution displays a sharp peak structure as if many Bose atoms 'condensate' at zero momentum. While with the increase of temperature the sharp peak disappear and TG gas distribute in high momentum regions.

The pair distribution and density profiles depend on the modulus of eigenfunction so they are same for TG gas and polarized free Fermions in the whole temperature region. The ROBDM and therefore the occupation number of natural orbital and momentum distribution depends on the eigenfunction so their properties for TG gas are different from those of free Fermions. But the distinction are shown only at low temperature, and we can not distinct them from each other qualitatively at high temperature.

\begin{acknowledgments}
This work is supported by NSF of China under Grants  No. 11004007, and ``the Fundamental Research Funds for the Central Universities."
\end{acknowledgments}


\begin{thebibliography}{99}

\bibitem{Ketterler}  N. J. van Druten and W. Ketterle, Phys. Rev. Lett. \textbf{79}, 549 (1997).
\bibitem{Paredes}  B. Paredes, A. Widera, V. Murg, O. Mandel, S. F\"{o}lling, I. Cirac, G. V. Shlyapnikov, T. W. H\"{a}nsch, and I. Bloch, Nature \textbf{429}, 277 (2004).
\bibitem{Toshiya}  T. Kinoshita, T. Wenger and D. S. Weiss, Science \textbf{ 305}, 1125 (2004).
\bibitem{Olshanii}  M. Olshanii, Phys. Rev. Lett. \textbf{81}, 938 (1998).
\bibitem{GirardeauJMP} M. D. Girardeau, J. Math. Phys. (N.Y.) \textbf {6}, 516 (1960).

\bibitem{Tonks} L. Tonks, Phys. Rev. \textbf{50}, 955 (1936).
\bibitem{Jacqmin} T. Jacqmin, J. Armijo, T. Berrada, K. V. Kheruntsyan, and I. Bouchoule, Phys. Rev. Lett. \textbf{106}, 230405 (2011).


\bibitem{LiebLiniger}  E. H. Lieb and W. Liniger, Phys. Rev. \textbf{130}, 1605 (1963).
\bibitem{CNYang} C. N. Yang and C. P. Yang, J. Math. Phys. \textbf{10}, 1115 (1969).

\bibitem{GaoXL} G. Xianlong, M. Polini, R. Asgari, and M. P. Tosi, Phys. Rev. A \textbf{73}, 033609 (2006); S. H. Abedinpour, M. Polini, G. Xianlong, and M. P. Tosi, Phys. Rev. A  \textbf{75}, 015602 (2007).
\bibitem{Hao09}  Y. Hao, and S. Chen, Phys. Rev. A. \textbf{80}, 043608 (2009).

\bibitem{Yin2009} X. Yin, S. Chen, and Y. Zhang,  Phys. Rev. A. \textbf{79}, 053604 (2009).

\bibitem{GuanXWPRL}    E. Zhao, X. W. Guan, W. V. Liu, M. T. Batchelor, and M. Oshikawa, Phys. Rev. Lett. \textbf{103}, 140404 (2009).

\bibitem{ALenard} A. Lenard, J. Math. Phys. 5, 930 (1964); A. Lenard, J. Math. Phys. (N. Y.)  \textbf{7} 1268 (1966).

\bibitem{Girardeau02}  M. D. Girardeau, Phys. Rev. Lett. \textbf{89}, 170404 (2002).
\bibitem{Zoellner}  S. Z\"{o}llner, H.-D. Meyer, and P. Schmelcher, Phys. Rev. A \textbf{74}, 063611 (2006).
\bibitem{Deuretzbacher}  F. Deuretzbacher, K. Bongs, K. Sengstock, and D. Pfannkuche, Phys. Rev. A \textbf{75}, 013614 (2007);  X. Yin, Y.Hao,  S. Chen and Y. Zhang, Phys. Rev. A. \textbf{78}, 013604 (2008).
\bibitem{Hao06}  Y. Hao, Y. Zhang, J. Q. Liang, and S. Chen, Phys. Rev. A. \textbf{73}, 063617 (2006).
\bibitem{Hao07}  Y. Hao, Y. Zhang, and S. Chen, Phys. Rev. A. \textbf{76}, 063601 (2007).
\bibitem{PDeuar} P. Deuar, A. G. Sykes, D. M. Gangardt, M. J. Davis, P. D. Drummond, and K. V. Kheruntsyan, Phys. Rev. A \textbf{79}, 043619 (2009).
\bibitem{ShlyapnikovPRL} K. V. Kheruntsyan, D. M. Gangardt, P. D. Drummond, and G. V. Shlyapnikov, Phys. Rev. Lett. \textbf{91}, 040403 (2003).
\bibitem{ShlyapnikovPRA} K. V. Kheruntsyan, D. M. Gangardt, P. D. Drummond, and G. V. Shlyapnikov, Phys. Rev. A \textbf{71}, 053615 (2005).



\bibitem{MRigol} M. Rigol, Phys. Rev. A. \textbf{72}, 063607 (2005).
\bibitem{JBrand}  A. Y. Cherny and Joachim Brand, Phys. Rev. A \textbf{73}, 023612 (2006).



\bibitem{MKormos} M. Kormos, G. Mussardo, and A. Trombettoni, Phys. Rev. A \textbf{81}, 043606 (2010).

\bibitem{HHu} H. Hu, X. L. Gao, and X. J. Liu, Phys. Rev. A \textbf{90}, 013622 (2014).

\bibitem{ShlyapnikovDisorder} I. L. Aleiner, B. L. Altshuler, and G. V. Shlyapnikov, Nat. Phys. \textbf{6}, 900 (2010).

\bibitem{AVogler} A. Vogler, R. Labouvie, F. Stubenrauch, G. Barontini, V. Guarrera, and H. Ott, Phys. Rev. A \textbf{88}, 031603 (2013).

\bibitem{MPanfil} M. Panfil and J. S. Caux, Phys. Rev. A \textbf{89}, 033605 (2014).

\bibitem{PVignolo} P. Vignolo and A. Minguzzi, Phys. Rev. Lett \textbf{110}, 020403 (2013).
\bibitem{MDHoffman} M. D. Hoffman, P. D. Javernick, A. C. Loheac, W. J. Porter, E. R. Anderson, and J. E. Drut, Phys. Rev. A \textbf{91}, 033618 (2015).
\bibitem{GLang} G. Lang, F. Hekking, and Anna Minguzzi, arxiv:1503.08038v1.

\bibitem{Girardeau2001} M. D. Girardeau, E. M. Wright, and J. M. Triscari,  Phys. Rev. A \textbf{63}, 033601 (2001).

\end{thebibliography}
\end{document}